\documentclass[aps,prd,preprint,superscriptaddress,showpacs]{revtex4}
\usepackage{epsfig}
\usepackage{dcolumn}
\usepackage{bm}
\usepackage{amsmath}
\usepackage{amsfonts}
\usepackage{amssymb}
\usepackage{graphicx}
\usepackage{latexsym}
\usepackage{multirow}
\usepackage{amssymb}
\usepackage{amsmath}
\usepackage{amsfonts}
\usepackage{mathrsfs}
\usepackage{ulem}
\usepackage{xcolor}
\usepackage{hyperref}

\begin{document}

\preprint{ACT-10-13, MIFPA-13-nn}

\title{Supergravity Inflation with Broken Shift Symmetry and Large Tensor-to-Scalar Ratio}

\author{Tianjun Li}
\affiliation{State Key Laboratory of Theoretical Physics
and Kavli Institute for Theoretical Physics China (KITPC),
      Institute of Theoretical Physics, Chinese Academy of Sciences,
Beijing 100190, P. R. China}

\affiliation{George P. and Cynthia W. Mitchell Institute for 
Fundamental Physics and Astronomy,
Texas A\&M University, College Station, TX 77843, USA}

\affiliation{School of Physical Electronics,
University of Electronic Science and Technology of China,
Chengdu 610054, P. R. China}

\author{Zhijin Li}

\affiliation{George P. and Cynthia W. Mitchell Institute for 
Fundamental Physics and Astronomy,
Texas A\&M University, College Station, TX 77843, USA}

\author{Dimitri V. Nanopoulos}

\affiliation{George P. and Cynthia W. Mitchell Institute for 
Fundamental Physics and Astronomy,
Texas A\&M University, College Station, TX 77843, USA}

\affiliation{Astroparticle Physics Group, Houston Advanced 
Research Center (HARC), Mitchell Campus, Woodlands, TX 77381, USA}

\affiliation{Academy of Athens, Division of Natural Sciences, 
28 Panepistimiou Avenue, Athens 10679, Greece}

\begin{abstract}

We propose a class of inflation models with potential 
$V(\phi)=\alpha \phi^n\rm{exp}(-\beta^m \phi^m)$. 
We show that such kind of inflaton potentials can be realized 
in supergravity theory with a small shift symmetry breaking 
term in the K\"ahler potential. We find that the models with 
($m=1,~ n=1$), ($m=1,~ n=2$),  ($m=2,~ n=1$), and ($m=2,~ n=3/2$)
can easily accommodate the Planck results. In particular, the 
tensor-to-scalar ratio is larger than $0.01$ in the $1\sigma$ region.
Thus, these models generate the typical large field inflation,
and can be tested at the future Planck and QUBIC experiments.

\end{abstract}

\maketitle

\newpage
\section{Introduction}



It is well-known that our Universe might experience an accelerated expansion, {\it i.e.},
inflation~\cite{Staro, Guth:1980zm, Linde:1981mu, Albrecht:1982wi}, at a very early stage of 
evolution, as suggested by the observed temperature 
fluctuations in the cosmic microwave background radiation (CMB).
Recently, the Planck satellite measured the CMB temperature anisotropy with 
an unprecedented accuracy. From its first-year observations, the scalar spectral index $n_s$, 
the running of the scalar spectral index $\alpha_s \equiv {\rm d} n_s/{\rm d~ln}k$,
the tensor-to-scalar ratio $r$, and the scalar amplitude $A_s$
for the power spectrum of the curvature perturbations  are respectively 
constrained to be~\cite{Ade:2013uln} 
\begin{eqnarray}
&& n_s = 0.9603 \pm 0.0073~,~~\alpha_s=-0.0134\pm 0.0090 ~,~~\nonumber \\
&& r \le 0.11~,~~A^{1/2}_s = 4.6856^{+0.0566}_{-0.0628} \times 10^{-5}~.~\,
\end{eqnarray}
Furthermore, there is no sign of primordial non-Gaussianity in the CMB fluctuations.
The generic predictions of the cosmological inflationary paradigm are
qualitatively consistent with the Planck results. However, 
a lot of previously popular inflation models are challenged. For
example, single field inflation models with a monomial potential $\phi^n$
for $n\ge 2$ are disfavoured due to large $r$. Interestingly, 
the Starobinsky $R+R^2$ model~\cite{Staro, MC}
predicts a value of $n_s \sim 0.96$, which is in perfect agreement 
with the CMB data, and a value of $r \sim 0.004$ that is comfortably 
consistent with the Planck upper 
bound~\cite{Ade:2013uln}.

From the particle physics point of view, supersymmetry is the most promising
extension for the Standard Model (SM). In particular,
the scalar masses can be stabilized, and the superpotential is 
non-renormalizable. Because gravity is also very important in the 
early Universe, it seems to us that supergravity theory is
a natural framework for inflation model building~\cite{SUGRA}. However,  
 supersymmetry breaking scalar masses in a generic supergravity theory
are of the same order as the gravitino mass, giving rise to 
the so-called $\eta$ problem~\cite{eta}, 
where all the scalar masses are of the order of the Hubble parameter
because of the large vacuum energy density during inflation~\cite{glv}.
There are two elegant solutions: no-scale supergravity~\cite{Cremmer:1983bf, 
Ellis:1984bf, Enqvist:1985yc, Ellis:2013xoa, Ellis:2013nxa, Li:2013moa, Ellis:2013nka},
and shift-symmetry in the K\"ahler potential~\cite{Kawasaki:2000yn, Yamaguchi:2000vm,
Yamaguchi:2001pw, Kawasaki:2001as, Kallosh:2010ug, Kallosh:2010xz, Nakayama:2013jka,
Nakayama:2013txa, Takahashi:2013cxa}. 

The Planck satellite  may measure the tensor-to-scalar ratio $r$ down to 
0.03-0.05 in one or two years. The future QUBIC experiment targets to constrain 
at the 90\% Confidence Level (C.L.) the tensor-to-scalar
ratio of 0.01 with one year of data taking from the Concordia
Station at ${\rm D\hat{o}me}$ C, Antarctica~\cite{Battistelli:2010aa}. 
Interestingly, the well-known Lyth bound
on tensor-to-scalar ratio is 0.01~\cite{Lyth:1996im}. 
Thus, the interesting question is how to construct inflation models which can 
be consistent with the Planck results and have large tensor-to-scalar ratio.
For recent studies, see 
Refs.~\cite{Li:2013moa, Nakayama:2013jka, Nakayama:2013txa, Croon:2013ana}.

According to the Planck observations, the simple chaotic inflation models based 
on power law potentials $\phi^n$ are out of $1\sigma$ region~\cite{Ade:2013uln}
due to the large tensor-to-scalar ratio. In this paper, we will modify
the simple power law potentials by multiplying an exponential term as follows
\begin{equation}
V(\phi)=\alpha \phi^n\rm{exp}(-\beta^m \phi^m)~,~\, 
\label{V}
\end{equation}
 where the parameter $\alpha$ relates to the energy scale of inflation, and $\beta$ reveals 
the importance of exponential term. For simplicity, we take the Planck
scale as unity $M_{\rm Pl}=1$ unless explicitly specified.
These potentials with a constraint $n=4(1-m)$ have been discussed in Ref.~\cite{R1} 
(For a review, see Ref.~\cite{Martin:2013tda}.), where the expansion scale factor 
$a(t)={\rm exp}(\alpha(\ln(t))^\beta)$ is adopted. Besides, a concrete example, 
$V(\phi)=\alpha \phi^2\rm{exp}(-\beta^2 \phi^2)$ with a non-canonical kinetic term 
for $\phi$ is 
obtained in ripple inflation from the modified no-scale supergravity~\cite{Enqvist:1985yc, Li:2013moa}. 
We show that the above inflaton potentials can be realized in the supergravity theory
with an small shift symmetry breaking term in K\"ahler potential. 
Because we do not consider the non-renormalizable terms in K\"ahler potential, 
there are two cases for $m$: $m=1$ and $m=2$. We find that 
 the models with ($m=1,~ n=1$), ($m=1,~ n=2$),  ($m=2,~ n=1$), and ($m=2,~ n=3/2$)
can be highly consistent with the Planck observations. Especially,  
the tensor-to-scalar ratio is larger than $0.01$ in the $1\sigma$ region, above the well-known 
Lyth bound~\cite{Lyth:1996im}. Thus, these models produce the typical large field inflation,
and can be tested at the future Planck and QUBIC experiments.

\section{Supergravity Inflation with Shift Symmetry Breaking}

The first step to realize inflation in supergravity is to get sufficient flat direction 
in the inflaton potential. The scalar potential in the supergravity theory with given 
K\"ahler potential $K$ and superpotential $W$ is
\begin{equation}
V=e^K\left((K^{-1})^{i}_{\bar{j}}D_i W D^{\bar{j}} \overline{W}-3|W|^2 \right)~,~
\label{sgp}
\end{equation}
where $(K^{-1})^{i}_{\bar{j}}$ is the inverse of the K\"ahler metric 
$K_{i}^{\bar{j}}=\partial^2 K/\partial \Phi^i\partial{\bar{\Phi}}_{\bar{j}}$, and $D_iW=W_i+K_iW$. 
For the canonical K\"ahler potential $K=\Phi\Phi^\dagger$, without miracle cancellation in Eq.~(\ref{sgp}), 
the scalar potential $V$ contains a term $e^{\Phi\Phi^\dagger}$. Because it is very steep in the region
with $\Phi>1$, 
 no slow-roll inflation can be realized for such potential. This is the so-called $\eta$ problem, 
the major difficulty to achieve inflation in supergravity. The $\eta$ problem can be naturally solved
in the no-scale supergravity~\cite{Cremmer:1983bf}, in which the K\"ahler potential satisfies the 
flatness condition $(K^{-1})^{i}_{\bar{j}} K_iK^{\bar{j}}=3$, and the scalar potential can be 
arranged to be sufficient flat for inflation.

For general supergravity with a polynomial K\"ahler potential, normally the scalar 
potential varies significantly 
above the Planck scale. Interestingly, with a certain symmetry, one can decouple the inflaton 
from the K\"ahler potential. So the inflaton does not appear in the $e^K$  term of 
the potential, and then no $\eta$ problem in this direction. This fact, which provides another solution 
to $\eta$ problem, realizes inflation in supergravity~\cite{Kawasaki:2000yn}. In such a case, 
the K\"ahler potential $K$ is invariant under the shift symmetry $\Phi\rightarrow\Phi+iC M_{\rm Pl}$
with $C$ a dimensionless real parameter, {\it i.e.}, 
the K\"ahler potential $K$ is a function of $\Phi+\Phi^{\dagger}$ and independent on the imaginary part 
of $\Phi$. Consequently, it forms a flat direction along the imaginary part of $\Phi$, and inflation 
may be triggered along this flat path.

Based on the fact that shift symmetry can keep the inflaton away from the $\eta$ problem, several inflation 
models in supergravity were 
proposed~\cite{Yamaguchi:2000vm, Yamaguchi:2001pw, Kawasaki:2001as, Kallosh:2010ug, Kallosh:2010xz}. 
As the inflation models with power law potentials are disfavored by the Planck observations, 
 the supergravity inflation models with polynomial 
potentials have been studied recently~\cite{Nakayama:2013jka, Nakayama:2013txa, Takahashi:2013cxa}. In those works, 
the K\"ahler potentials are constrained by the shift symmetry in certain direction, while 
the superpotentials are generalized to be polynomial potentials.

In the following, we will obtain the potential $\alpha\phi^ne^{-\beta^m\phi^m}$ in supergravity 
by introducing a new term in K\"ahler potential which slightly breaks the shift symmetry. 
Because the mass dimension of K\"ahler potential
is two, we will not consider the models with $m\ge 3$ in this paper. 
The detailed studies for the inflation
models with spontaneous shift symmetry breaking and the inflation
models with $m\ge 3$ will be given elsewhere~\cite{LLN-Prep}. 

Before we study the general cases, a special potential with ($m=1$, $n=2$) will be
considered as a concrete example.

\subsection{Supergravity Realization of Potential $\alpha\phi^ne^{-\beta\phi}$}

The K\"ahler potential and superpotential used to get the potential $\alpha\phi^2e^{-\beta\phi}$ are
\begin{equation}
\begin{split}
K=-b(\Phi+\Phi^\dagger)-\frac{1}{2}(\Phi-\Phi^\dagger)^2+XX^\dagger~,\\
W=a\Phi X~,~~~~~~~~~~~~~~~~~~~
\end{split} \label{kah1}
\end{equation}
in which the term $-\frac{1}{2}(\Phi-\Phi^\dagger)^2$ has the shift symmetry $\Phi\rightarrow \Phi+C M_{\rm Pl}$. 
This symmetry is broken by the linear term $-b(\Phi+\Phi^\dagger)$. As shown later, the
shift symmetry breaking term 
introduces an exponential term $e^{-\beta\phi}$ in the potential, which is crucial to fit
the Planck results. Actually, the model in Eq.~(\ref{kah1}) is equivalent to the supergravity theory
with canonical K\"ahler potential and superpotential as follows
 \begin{equation}
\begin{split}
K=\Phi\Phi^\dagger+XX^\dagger~, ~~\\
W=ae^{-b\Phi-\frac{1}{2}\Phi^2}\Phi X~.~\,
 \label{kah-A}
 \end{split}
 \end{equation}
The new superpotential leads to the ``miracle cancellation", which plays the same role 
as the shift symmetry of K\"ahler potential in Eq.~(\ref{kah1}) to solve the $\eta$ problem.
In addition, our model is equivalent to the supergravity model in 
Refs.~\cite{Kallosh:2010ug, Kallosh:2010xz} where the K\"ahler potential 
and superpotential are
\begin{equation}
\begin{split}
K=-\frac{1}{2}(\Phi-\Phi^\dagger)^2+XX^\dagger~,\\
W=ae^{-b\Phi}\Phi X~.~\,
\end{split} \label{kah-B}
\end{equation}
Especially, the shift symmetry is preserved in the K\"ahler potential. As we know, the
 superpotentials in Eqs.~(\ref{kah-A}) and (\ref{kah-B}) 
are not the traditional potentials in 
particle physics and string theory, and can not be generated by 
instanton effect or strong dynamics, etc. Thus, we do not consider these
alternative explanations in this paper.

The scalar potential is
\begin{eqnarray}
V&=&a^2{\rm exp}\left(-b(\Phi+\Phi^\dagger)-\frac{1}{2}(\Phi-\Phi^\dagger)^2+XX^\dagger \right) \nonumber \\
&& \left(\Phi\Phi^\dagger+XX^\dagger-bXX^\dagger(\Phi+\Phi^\dagger)+(1+b^2)\Phi\Phi^\dagger XX^\dagger 
\right.\nonumber\\&&\left.
- XX^\dagger(\Phi^2+{\Phi^\dagger}^2)-
\Phi\Phi^\dagger XX^\dagger (\Phi-\Phi^\dagger)^2+\Phi\Phi^\dagger (XX^\dagger)^2 \right)~.~\,
\label{potential1}
\end{eqnarray}
Decomposing the complex field $\Phi$ as follows
\begin{equation}
\Phi=\frac{1}{\sqrt{2}}(\phi+i\chi)~,~
\end{equation}
we obtain the scalar potential 
\begin{eqnarray}
V&=&a^2e^{-\sqrt{2}b\phi+\chi^2+XX^\dagger} \left(\frac{1}{2}(\phi^2+\chi^2)+XX^\dagger
-\sqrt{2}b\phi XX^\dagger+\frac{b^2-1}{2}\phi^2 XX^\dagger  
 \right.\nonumber\\&&\left.
+\frac{b^2+3}{2}\chi^2 XX^\dagger + \chi^2(\phi^2+\chi^2)XX^\dagger +\frac{1}{2}(\phi^2+\chi^2) 
(XX^\dagger)^2 \right)~.~
\label{potential2}
\end{eqnarray}
So the potential $V$ depends on $\chi$ and $X$ through $e^{\chi\chi^\dagger}$ and $e^{XX^\dagger}$, 
as expected. While the shift symmetry breaking term $-b(\Phi+\Phi^\dagger)$ introduces a new term 
$e^{-\sqrt{2}b\phi}$ in the potential, which makes the potential flatter in the inflaton $\phi$
direction.

The masses of $\chi$ and $X$ with non-zero $\phi$ satisfy the following conditions
\begin{equation}
\begin{split}
m_\chi^2=2V+a^2e^{-\sqrt{2}b\phi+\chi^2+XX^\dagger} \left(1+(b^2+3)XX^\dagger+\cdots \right)>6H^2~,~\, \\
m_X^2=V+a^2e^{-\sqrt{2}b\phi+\chi^2+XX^\dagger} \left(1-\sqrt{2}b\phi+\frac{b^2-1}{2}\phi^2
+2\phi^2XX^\dagger+\cdots \right)~,~\,
\end{split} \label{masses}
\end{equation}
where $H^2 = V/3$. Because $b$ is a small number for shift symmetry breaking, 
$m^2_X$ will be smaller than $3 H^2$ if the magnitude of $X$ is smaller than $1/2$.
To avoid having a very light scalar field $X$ during
inflation, we can introduce a quartic term $-\xi (XX^{\dagger})^2$ in the
K\"ahler potential~\cite{Kallosh:2010ug, Kallosh:2010xz}.
Therefore, both $\chi$ and $X$ vanish fast and have little effects on 
the slow-roll inflation process. 
They can be safely fixed at the origin during inflation. The inflaton scalar
potential in Eq.~(\ref{potential2}) 
with $\chi=X=0$ is simplified as follows
\begin{equation}
V=\frac{1}{2}a^2\phi^2e^{-\sqrt{2}b\phi},
\end{equation}
which is the same as the potential proposed in Eq.~(\ref{V}) with ($m=1$, $n=2$) 
under the relations $\alpha=\frac{1}{2}a^2$ and $\beta=\sqrt{2}b$.

To get the general potential $\alpha \phi^n\rm{exp}(-\beta \phi)$, we only need to use 
the same K\"ahler potential as above and a new superpotential 
\begin{equation}
W=a\Phi^{\frac{n}{2}}X ~.~\,
\label{Superpotential}
\end{equation}
 It is easily 
shown that the extra fields $\chi$ and $X$ are still fixed at the origin during inflation, 
and then the inflaton scalar potential is
\begin{equation}
V=e^{-b(\Phi+\Phi^\dagger)}|W_X|^2=2^{-\frac{n}{2}}a^2\phi^ne^{-\sqrt{2}b\phi}~.~\,
\end{equation}

\subsection{Supergravity Realization of Potential $\alpha\phi^ne^{-\beta^2\phi^2}$}

For the potentials with $m=2$, we employ a different K\"ahler potential which has 
quadratic shift symmetry breaking term
\begin{equation}
K=-b^2(\Phi^2+\Phi^{\dagger 2})-\frac{1}{2}(\Phi-\Phi^\dagger)^2+XX^\dagger~,~\,   
\label{kah2}
\end{equation}
and the superpotential given by Eq.~(\ref{Superpotential}). 
Similar to the K\"ahler potential 
in Eq.~(\ref{kah1}), the above K\"ahler potential possesses the shift symmetry 
$\Phi\rightarrow \Phi+C M_{\rm Pl}$ if $b=0$, while this shift 
symmetry is slightly broken for a small but non-zero $b$.

For the K\"ahler potential in Eq.~(\ref{kah2}), taking the complex field 
$\Phi=\frac{1}{\sqrt{2}}(\phi+i\chi)$, 
we obtain the exponential term in scalar potential
\begin{equation}
V\propto e^{-b^2\phi^2+(1+b^2)\chi^2+XX^\dagger}.
\end{equation}
All the other components in the potential are identical to the above scenario with  $m=1$. 
The mass of complex field $X$ is 
the same as that in Eq.~(\ref{masses}), while the mass of the real field $\chi$ is even larger
\begin{equation}
{m_\chi}^2=2(1+b^2)V+\cdots>6(1+b^2)H^2~.~
\end{equation}
Consequently, both $\chi$ and $X$ vanish during inflation. 
And the inflaton scalar potential is
\begin{equation}
V=e^{-b^2\phi^2}|W_X|^2=2^{-\frac{n}{2}} a^2\phi^ne^{-b^2\phi^2}~.
\end{equation}
Therefore, the inflation potential with $m=2$ can also be realized in supegravity 
by introducing the shift symmetry breaking term in K\"ahler potential.

\section{Slow-Roll Inflation}

Based on the Planck results, inflation models with the pure power law potentials are disfavored: 
the $n_s-r$ relations with the e-folding number range $N\in [50, 60]$ are out of the $1\sigma$ region
due to large $r$. 
With the exponential terms in the potentials, the $n_s-r$ relation will be significantly improved
because of the following two points: (1) The potentials will become flatter due to exponential terms 
and more suitable for inflation; 
(2) The exponential term introduces one more free parameter $\beta$, which can be 
adjusted to fit the observation data.

\begin{figure}
\centering
\includegraphics[width=80mm, height=49mm,angle=0]{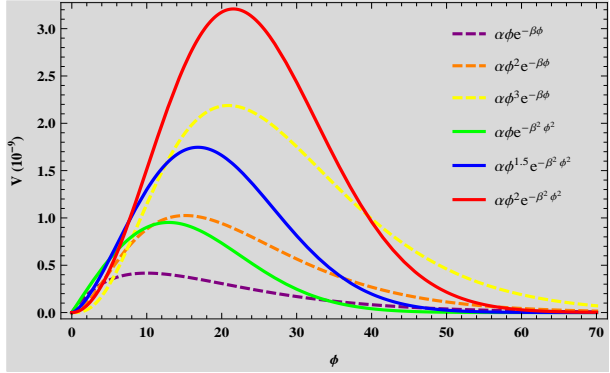}
\caption{The potentials  $\alpha \phi^n\rm{exp}(-\beta \phi)$ and $\alpha \phi^n\rm{exp}(-\beta^2 \phi^2)$
for various $n$. 
Here, the parameters $\alpha$ and $\beta$ are chosen to 
be consistent with the Planck observations, as shown later.}
\label{potential}
\end{figure}

The inflaton scalar potentials in Eq.~(\ref{V}) with $m=1$ and $m=2$ are presented
in Fig.~\ref{potential} for various $n$. The inflation processes are
 assumed to be generated when the inflaton 
evolves from the local maxima to the origin.

\subsection{Inflation Models with $m=1$}

Taking $m=1$, the inflaton potential turns into
\begin{equation}
V(\phi)=\alpha \phi^n\rm{exp}(-\beta \phi). \label{V1}
\end{equation}
The inflation path is between the origin and the turning point $\phi_t=\frac{n}{\beta}$. The slow-roll 
parameters $\epsilon$ and $\eta$ are
\begin{equation}
\begin{split}
\epsilon=\frac{(n-\beta\phi)^2}{2\phi^2}~, ~~~~~~~~~~\\
\eta=\frac{(n-\beta\phi)^2-n}{\phi^2}=2\epsilon-\frac{n}{\phi^2}~.~\,
\end{split}
\end{equation}
During the inflation, the following slow-roll conditions should be satisfied
\begin{equation}
\epsilon\ll1~, ~~~~|\eta|\ll1~.
\end{equation}
At the end of inflation, we have either $\epsilon=1$ or $|\eta|=1$. In fact, 
the condition $\epsilon\ll1$ will be violated first for small $n\leqslant2$. From 
the condition $\epsilon=1$, the exit point of inflation is determined as 
$\phi_f=n/(\sqrt{2}+\beta)$, and the corresponding $\eta$ is
\begin{equation}
 \eta=2-\frac{(\sqrt{2}+\beta)^2}{n} ~.~\,
\label{eta}
\end{equation}
As we will see later, $\beta$ is very small to realize inflation processes that 
are consistent with the Planck observations. 
So for small $n\leqslant2$, $\eta$ is always smaller than $1$ before the inflation exit point. 
Its minimal value between $[\phi_f, \phi_t]$ is $-\beta^2-2\sqrt{2}\beta $ for $(n=1)$  or 
$-\frac{\beta^2}{n-1}$ for $(n>1)$, therefore, the slow-roll condition $|\eta|<1$ is satisfied. 
While for $n=3$,  the condition $|\eta| \ll1$ will be violated first, 
then the exit point of inflation $\phi_f$ is determined by the second slow-roll violation
 condition $|\eta|=1$
\begin{equation}
\phi_f=\frac{1}{1-\beta^2}(-3\beta+\sqrt{6+3\beta^2})~.
\end{equation}
The e-folding number $N$ is
\begin{equation}
N=\int_{\phi_i}^{\phi_f}d\phi ~\frac{V}{V_\phi}=-\frac{\phi}{\beta}-\frac{n}{\beta^2}\rm{ln}(n-\beta \phi)\big|_{\phi_i}^{\phi_f}~.
\end{equation}

\subsection{Inflation Models with $m=2$}

The inflaton potential with $m=2$ is
\begin{equation}
V(\phi)=\alpha \phi^n\rm{exp}(-\beta^2 \phi^2)~.
\end{equation}
The turning point of this potential is $\phi_t=\sqrt{\frac{n}{2}}\frac{1}{\beta}$, 
and the slow-roll parameters $\epsilon$ and $\eta$ are
\begin{equation}
\begin{split}
\epsilon=\frac{(n-2\beta^2\phi^2)^2}{2\phi^2}~,~~~~~~~~~~~~~~&
\\
\eta=-2\beta^2(1+2n)+\frac{n(n-1)}{\phi^2}+4\beta^4\phi^2\\
=2\epsilon-2\beta^2-\frac{n}{\phi^2}~.~~~~~~~~~~~~~~
\end{split}
\end{equation}
The condition $\epsilon=1$ or $|\eta|=1$ determines the exit point of inflation. The first condition 
will be fulfilled at point $\phi_f=\frac{1}{2\sqrt{2}\beta^2}(-1+\sqrt{1+4n\beta^2})$, and 
the corresponding $\eta$ is
\begin{equation}
\eta=2-\frac{1}{n} \left(1+\sqrt{1+4n\beta^2} \right)~.
\end{equation}
So for $n\leqslant2$, $\eta<1$ at the exit point $\phi_f$. 
Besides, it is easy to check 
that in the range $\phi\in [\phi_f, \phi_t]$, the minimal value of $\eta$ is larger than $-1$.
Thus, for the inflation models with $m=2$ and $n\leqslant2$, $\phi_f$ as given above is the exit point 
of inflation processes. The e-folding number $N$ is
\begin{equation}
N=\frac{1}{4\beta^2} \rm{ln}\frac{n-2\beta^2\phi_f^2}{n-2\beta^2\phi_i^2}~.~\,
\end{equation}

\section{Numerical Results}

\begin{figure}
\centering
\includegraphics[width=60mm, height=48mm,angle=0]{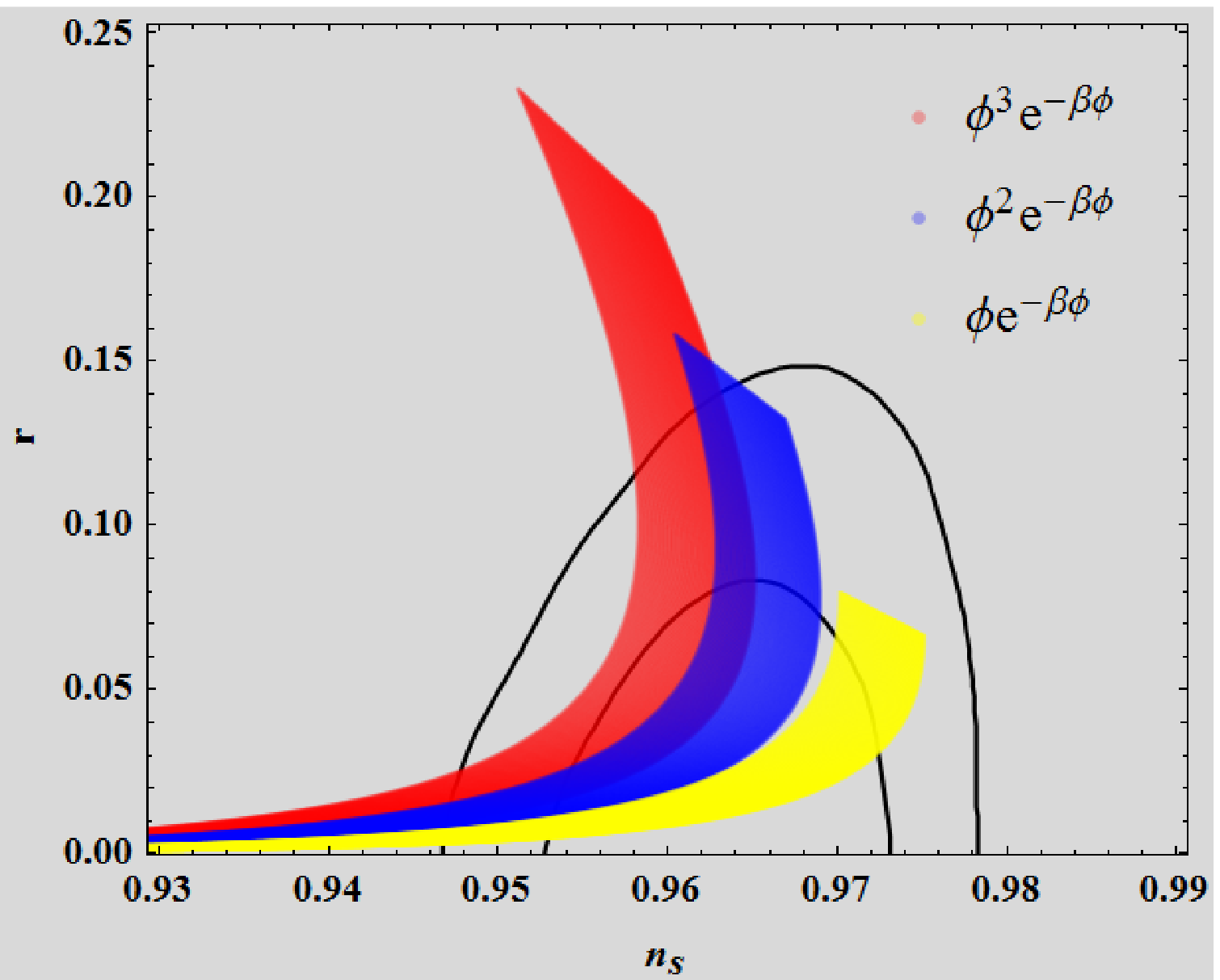}
\includegraphics[width=60mm, height=48mm,angle=0]{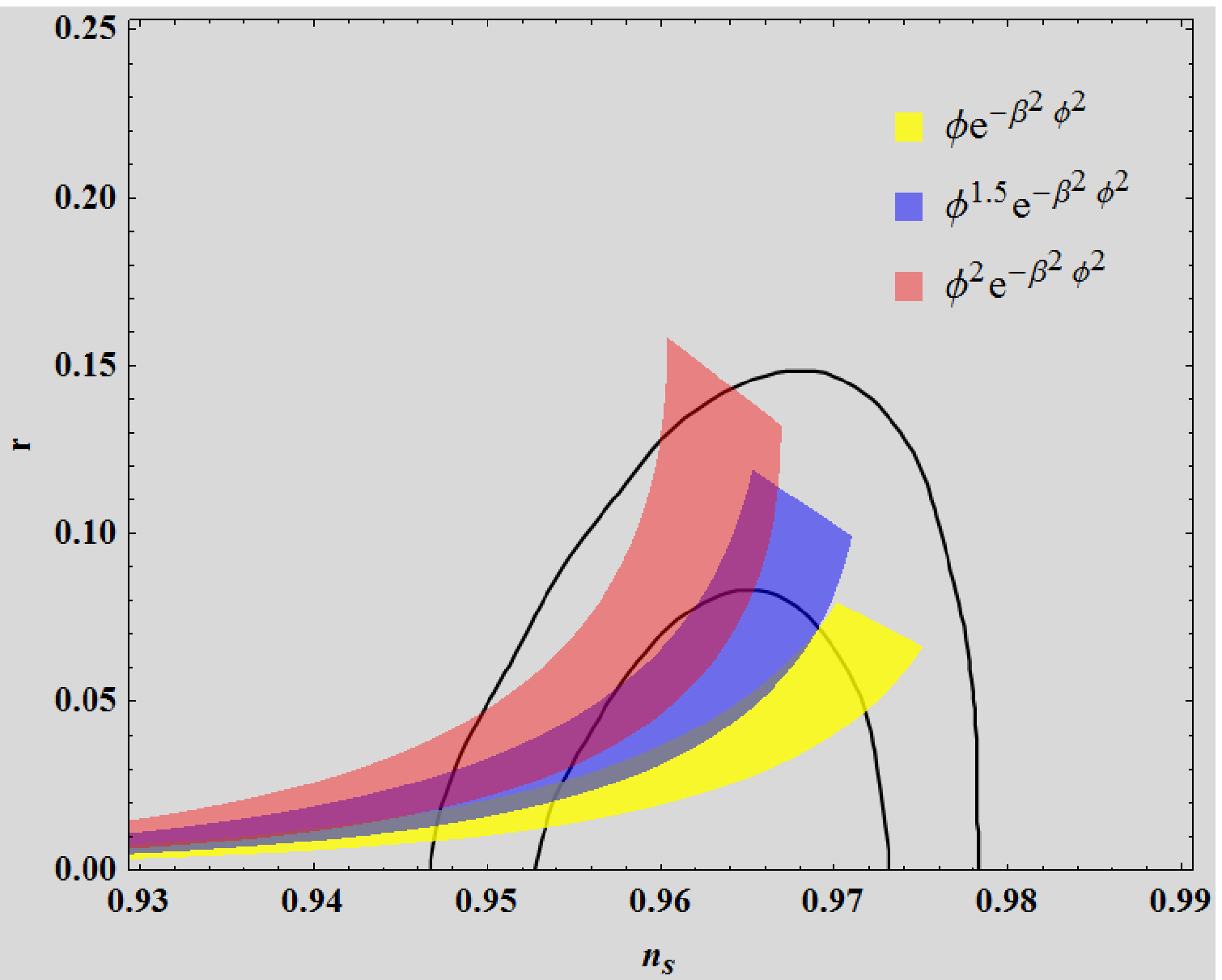}
\caption{$r$ versus $n_s$ for $\phi^n e^{-\beta \phi}$(left) and
  $\phi^n e^{-\beta^2 \phi^2}$(right). For each strip, the upper boundary corresponds to 
the constraint on the e-folding number $N=50$ while the lower boundary relates to $N=60$. The two dark lines 
provide the $1\sigma$ and $2\sigma$ regions from the Planck observations. 
For each strip, it ends at a certain line 
which corresponds to the prediction of the model with purely power law potential $\phi^n$.} 
\label{ns-r}
\end{figure}

The $n_s-r$ relations for the inflation models with $m=1$ and $2$ are given in Fig.~\ref{ns-r}. 
In particular, 
for the models with ($m=1,~ n=1$), ($m=1,~ n=2$),  ($m=2,~ n=1$), and ($m=2,~ n=3/2$), the predictions 
highly agree with the Planck observations. Notably, 
the tensor-to-scalar ratio is larger than $0.01$ in the $1\sigma$ region, above the well-known 
Lyth bound~\cite{Lyth:1996im}. So these models generate the typical large field inflation. 
In the following years the Planck experiment may measure the tensor-to-scalar ratio down to 0.03-0.05, 
and the future QUBIC experiment may measure $r$ down to 0.01. Thus, 
it would be very interesting to compare the future observations with the predictions of our models.

For $m=1$ and $m=2$, the models with smaller $n$ are preferred. As can be seen from the two graphs 
in Fig.~\ref{ns-r}, the strips have a tendency to run out of the $1\sigma$ region for larger $n$. 
For $m=1$ and $m=2$, the models respectively with $n>4$ and $n>3$ are 
out of the $1\sigma$ region.  So to be consistent with 
the observation data, $n$ cannot be too large.  The strips for the number of e-folding $N$ 
from 50 to 60 end at certain lines, 
which correspond to the predictions of the chaotic inflation models with potential $\phi^n$ 
(See Fig.~1 in Ref.~\cite{Ade:2013uln}.). In our models, when 
the parameter $\beta$ is close to zero, the exponential terms are ignorable. Thus,
we return to the classical chaotic inflation models with potential $\phi^n$, which 
gives the upper limits for the predictions.

\begin{figure}
\centering
\includegraphics[width=60mm, height=40mm,angle=0]{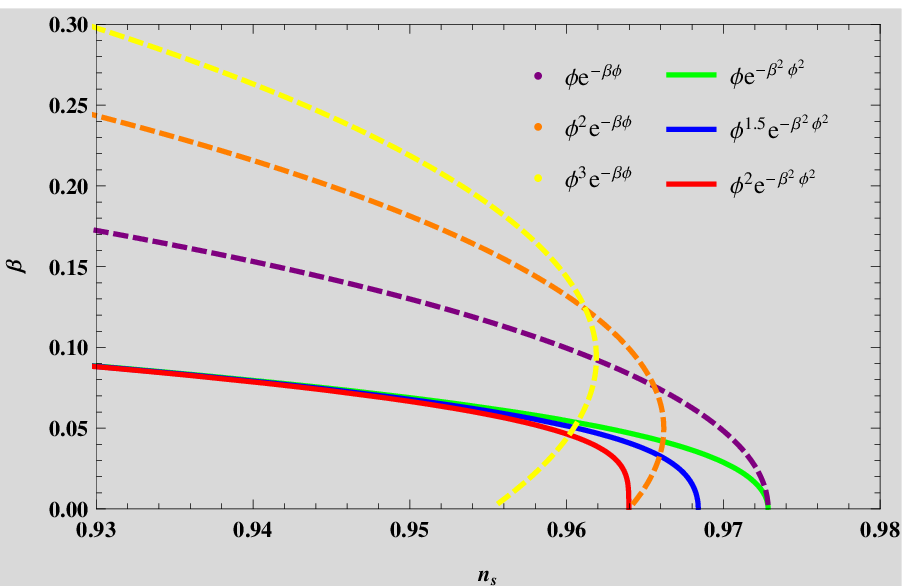}
\includegraphics[width=60mm, height=40mm,angle=0]{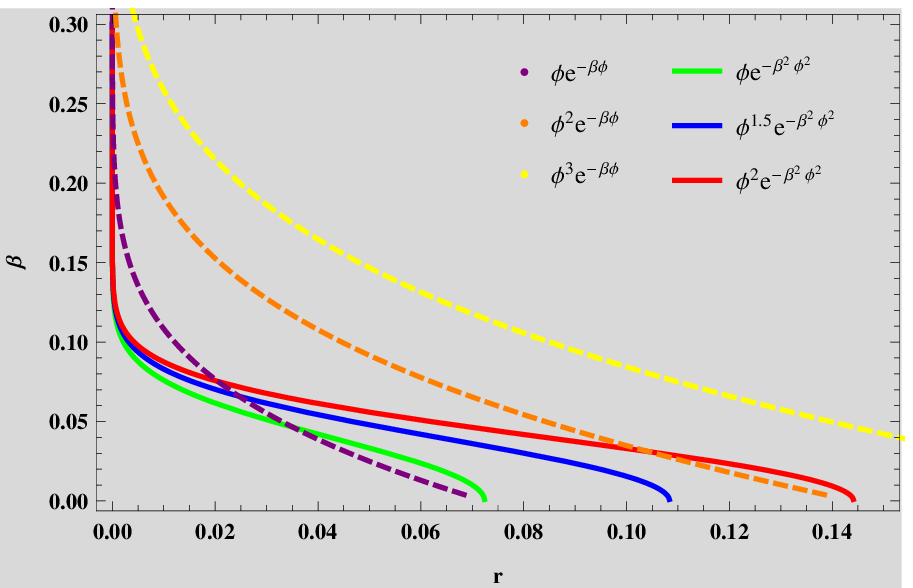}
\caption{$\beta$ versus $n_s$ (left) and
$\beta$ versus $r$  (right). Each line is drawn with the e-folding number $N=55$.} \label{nsr-beta}
\end{figure}

The relations between the parameter $\beta$ and inflation observables $n_s$/$r$
for $N=55$ are given
in Fig.~\ref{nsr-beta}. In general, to agree with the Planck observations 
in the $1\sigma$ region, the parameter $\beta$ 
should be small ($<0.25$), but fortunately, $\beta$ is just close to the order of $O(10^{-1})$, 
no fine-tuning is needed here to fit the observation data. The $\beta-r$ relations  
in Fig.~\ref{nsr-beta} 
are very interesting: to realize large $r$, $\beta$ should be very small, {\it i.e.}, the exponential terms 
become less and less important when $r$ increases; In contrast, for the inflation with small $r$, 
the effects of the exponential terms becomes more important and non-ignorable.

Besides the $\phi^n$ models, the chaotic inflation can also be realized by the polynomial 
potentials~\cite{Martin:2013tda}. Soon after the Planck results, the chaotic inflation based 
on the polynomial potential was studied~\cite{Croon:2013ana}, which can have large
$r$ as well. The potential is 
$a\phi^2(b-\phi)^2$ and the predictions are consistent with the Planck observations. 
The results in Ref.~\cite{Croon:2013ana} are similar to our model with $\phi^2e^{-\beta\phi}$ shown 
in Fig.~\ref{ns-r}. Actually, their potential can be considered as the approximation of our
 models with ($m=1, ~n=2$) 
\begin{equation}
V(\phi)=\alpha\phi^2e^{-\beta\phi}\simeq\alpha\phi^2(1-\beta\phi+\frac{1}{2}\beta^2\phi^2-\cdots)~,
\end{equation}
where $\beta$ is small and then all the non-renormalizable terms can be neglected. 
In other words, this approximation is valid for large $r$  and becomes worse for small $r$.

\begin{figure}
\centering
\includegraphics[width=60mm, height=40mm,angle=0]{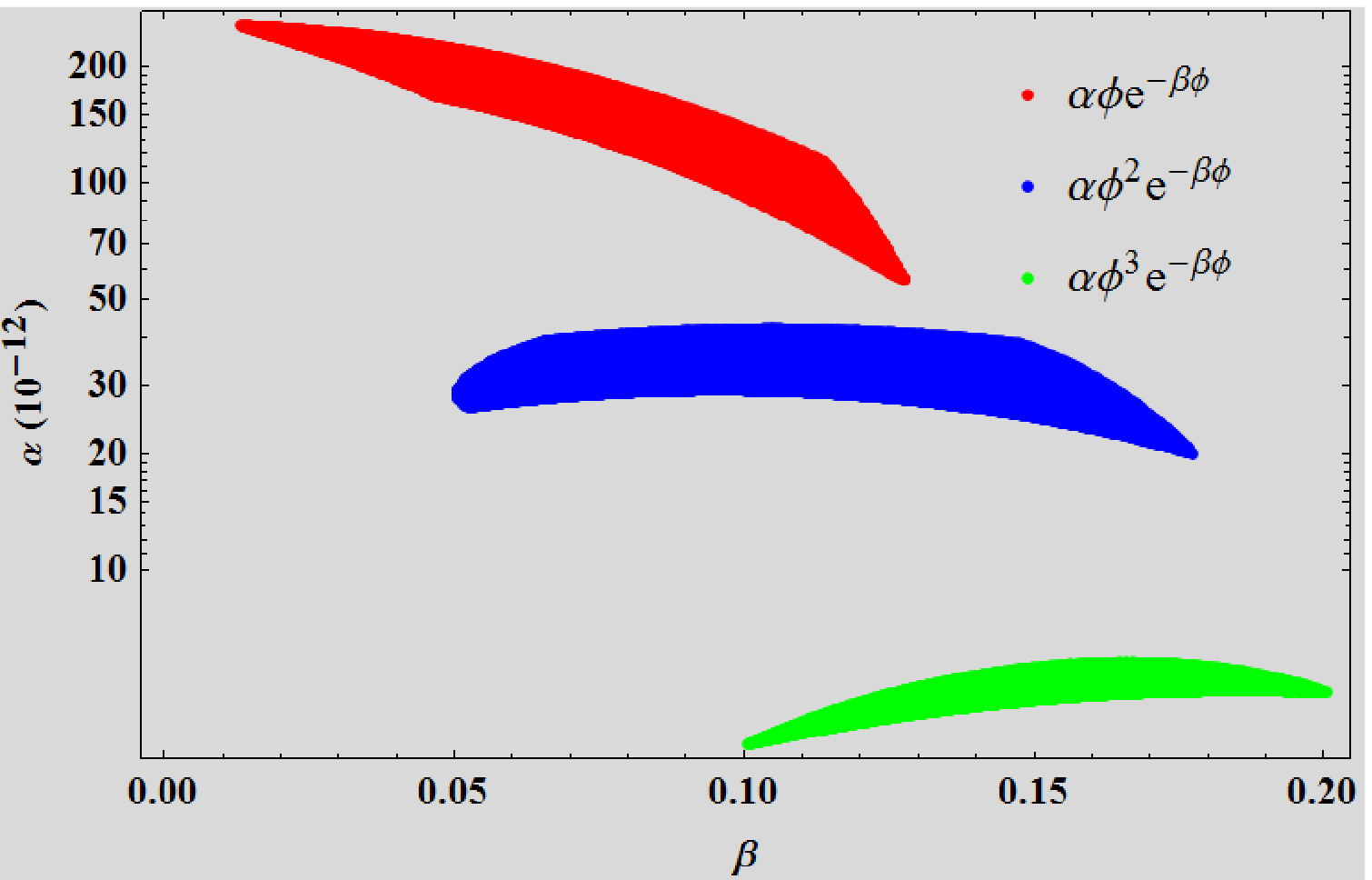}
\includegraphics[width=60mm, height=40mm,angle=0]{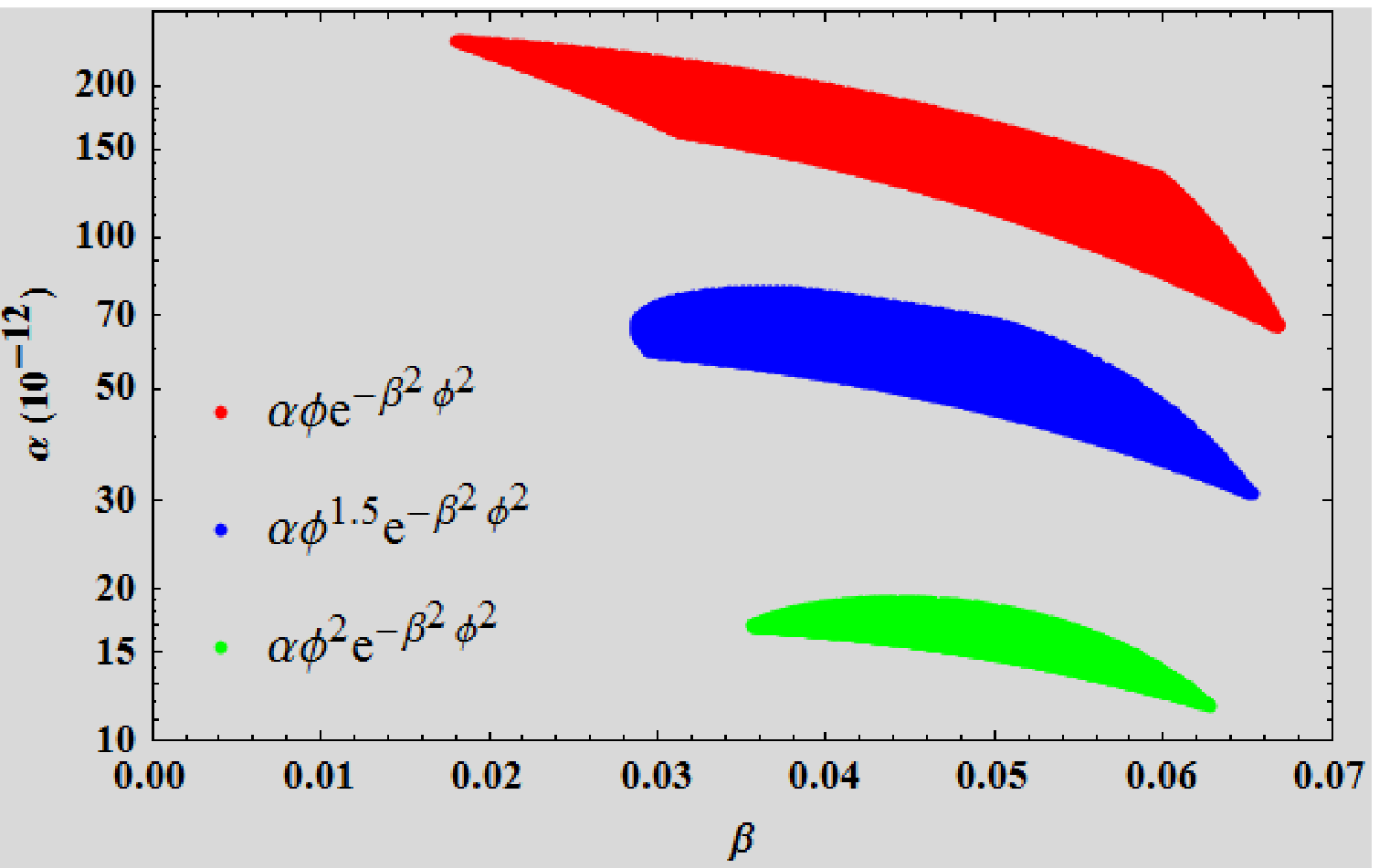}
\caption{Regions for $\alpha$ and $\beta$ in the inflation models $\alpha \phi^n e^{-\beta\phi}$ (left panel) 
and $\alpha \phi^n e^{-\beta^2\phi^2}$ (right panel) that are consistent with 
the Planck observations within $1\sigma$ region. 
Here, $A_s^{\frac{1}{2}}$ is fixed to be $4.686\times 10^{-5}$, and the effect of its uncertainty on 
the parameter $\alpha$ 
can be easily seen as they linearly depend on each other.} 
\label{pars}
\end{figure}

The second parameter  $\alpha$ in the model determines the scale of inflation potential, 
which directly relates to the scalar amplitude $A_s$
for the power spectrum of the curvature perturbation 
\begin{equation}
\alpha=\frac{3}{2}\pi^2 r A_s \phi_i^{-n}e^{\beta^m\phi_i^m}~, \label{As}
\end{equation}
where  $\phi_i$ is the inflaton when the mode $k_*$ crosses the Hubble radius for
the first time,  and the parameter $A_s$, according to 
the Planck observation, is about $2.196\times10^{-9}$. From Eq.~(\ref{As}), we can determine 
the regions for two parameters $\alpha$ and $\beta$ which agree with the Planck observations 
in the $1\sigma$ region. The results are given in Fig.~\ref{pars}. The parameter $\beta$ 
is of the order $10^{-1}$ or $10^{-2}$, which is natural as pointed out before. 
Another parameter $\alpha$ is 
of the order $10^{-10}$ for $n=1$, and gets even smaller for the models with $n>1$.
In the supergravity inflation models, we have $\alpha~=~2^{-n/2}a^2$.
Thus, $a$ is of the order $10^{-5}$, and then it is not very fine-tuned for inflation.

\section{Conclusion}


We have studied a class of inflation models with potential 
$V(\phi)=\alpha \phi^n\rm{exp}(-\beta^m \phi^m)$, which can 
be realized in the supergravity theory with an small shift 
symmetry breaking term in the K\"ahler potential. We showed that 
the models with ($m=1,~ n=1$), ($m=1,~ n=2$),  
($m=2,~ n=1$), and ($m=2,~ n=3/2$) have very good agreement
with the Planck observations. Especially,  
the tensor-to-scalar ratio is larger than $0.01$ in the $1\sigma$ region.
Therefore, our models realize the typical large field inflation,
and can be tested at the future Planck and QUBIC experiments.


\begin{acknowledgments}

This research was supported in part
by the Natural Science Foundation of China
under grant numbers 10821504, 11075194, 11135003, and 11275246, 
and by the DOE grant DE-FG03-95-Er-40917.

\end{acknowledgments}

\end{document}